# The origin of soft X-rays in luminous AGN

**Karl Mannheim**[1]**, Marcus Schulte**[1]**, and Jörg Rachen**[2]

[1] Universitäts-Sternwarte, Geismarlandstr. 11, D-37083 Göttingen, Germany
[2] Max-Planck-Institut für Radioastronomie, Auf dem Hügel 69, D-53121 Bonn, Germany

**Abstract.** We examine the effects of a nuclear jet expelled from an accretion disk on the UV-to-X-ray spectra of quasars. The base of the jet is immersed in the UV radiation field from the disk and heats up to the Compton-temperature scattering the UV photons from the disk into the soft X-ray range. *Unsaturated comptonization* leads to a power law extension of the UV bump spectrum to higher frequencies with spectral index $\alpha_s \gtrsim 1.7$ ($S_\nu \propto \nu^{-\alpha_s}$). In the keV range, a nonthermal hard X-ray component with flatter spectral index shows up. This emission component is assumed to be associated with *nonthermal* processes in the jet. In radio-louds, where the jet is highly collimated and relativistic, particle acceleration occurs at a large distance from the disk, producing a boosted $\alpha_h \sim 0.5$ hard X-ray spectrum at a low radiation compactness. In radio-quiets, the jet is slow and highly dissipative a few scale heights above the disk, producing hard X-rays with spectral index $\alpha_h \sim 1$ by pair cascades at a moderate radiation compactness. The jet provides the energetical link between the disk and the hard X-ray source. Modeling the broad-band spectra of several quasars, we find that soft X-ray excesses in the keV range occur when the kinetic power of the jet is of the same order as the UV luminosity.

## 1. Introduction

It is now well-established that the soft X-ray spectra of AGN show a more or less pronounced soft X-ray excess above a hard X-ray power law spectrum over a broad range of luminosities (e.g., Walter & Fink 1993, Saxton et al. 1993, Walter et al. 1994, Fiore et al. 1994). The soft X-ray emission component seems to vary independently of the hard X-ray emission component (Turner et al. 1990), and the steepness of the soft X-ray component suggests its connection with the ultraviolet

*Send offprint requests to*: K. Mannheim (kmannhe@uni-sw.gwdg.de)

bump. In some low-luminosity AGN, the ultrasoft X-ray excess seems to represent the Wien-tail of a thermal UV-to-soft X-ray bump, possibly due to an accretion disk (Grupe et al. 1995). In the luminous quasars, a Wien-spectrum reaching up to the soft X-ray range is conceptually difficult to reconcile with the accretion disk scenario, since the effective temperature of disks around supermassive black holes with masses $\sim 10^9 M_\odot$ is only of order $\sim 10$ eV (Laor & Netzer 1989). Accretion at a near-Eddington rate can lead to a comptonizing corona shifting the high-frequency tail of the thermal bump spectrum towards the soft X-ray range (Czerny & Elvis 1987, Ross et al. 1992). Apparently, near-Eddington accretion is in contradiction to the results of multifrequency modeling which generally yields sub-Eddington accretion rates (Sun & Malkan 1989). Moreover, pure $\alpha$-accretion-disk models explain neither hard X-ray emission nor radio jets, which are both frequently found in quasars. Another hint is that the EUV-to-soft X-ray spectrum in luminous sources seems to be a power law rather than an exponential (Puchnarewicz et al. 1994, Wisotzki et al. 1995, Fiore et al. 1994, Rachen et al. 1995). Thermal reprocessing of soft X-rays, albeit present at some level, generally still requires intrinsic soft X-ray spectra steeper than the hard X-ray spectra (e.g., Turner et al. 1993, Pounds et al. 1994). At high X-ray energies, thermal reprocessing gives rise to fluorescent iron lines and a Compton-reflection hump. The slope of the primary hard X-ray spectrum is inferred to be $\alpha_h \sim 0.9$, whereas the superposition of reflected and direct continuum leads to an apparent slope $\alpha_h \sim 0.7$ – the canonical X-ray index for Seyfert galaxies. In *quasars*, there is less evidence for reprocessing and the canonical index is $\alpha_h \sim 1$ for radio-quiets and $\alpha_h \sim 0.5$ for radio-louds (Wilkes & Elvis 1989). The primary continuum slope $\alpha_h \sim 0.9$-$1.0$ can be robustly produced by electromagnetic cascades (Svensson 1987). In relativistic jets, cascades can produce a boosted hard X-ray spectrum with $\alpha_h \sim 0.5$ (Mannheim 1993, Blandford & Levinson 1995). Recent detections of blazars in gamma-rays (Montigny et al. 1995) suggest that electromagnetic cascades are important as a radiation mechanism, indeed. However, radio-quiets are not detected in gamma-rays, and their X-ray spectra have been observed to turn over steeply at $\leq 100$ keV (Maisack et al. 1993,1995). Compton-upscattering of photons on thermal electrons rather than downscattering of photons in an electromagnetic cascade therefore seems to be favored in radio-quiets (Haardt & Maraschi 1991, Zdziarski et al. 1994). Energetically, such an X-ray spectrum can not produce the ultraviolet bump through reprocessing by optically thick matter (Lightman & White 1988, Guilbert & Rees 1988). Alternatively, the primary X-ray continuum in radio-quiets could still be nonthermal if it is anisotropic with most flux emitted towards the disk (Ghisellini et al. 1991, Mannheim 1995).

**Table 1.** Parameters used to obtain the multifrequency data fits shown in Fig. 1. The electron temperature was held fixed at $T_e = 5 \cdot 10^7$ K. The *ROSAT* PSPC fits yielding $\alpha_s$, $\alpha_h$ and $\epsilon_b$ were obtained for Galactic $n_H$, unless indicated otherwise. $L^h_{17.7}/L_{15.5}$ denotes the monochromatic luminosity ratio between the 2 keV luminosity *of the hard X-ray component* obtained from our spectral fits and the UV luminosity at $\log[\nu/\text{Hz}] = 15.5$. Mean values were computed for the fits indicated by a black dot. For a discussion of PHL 1657, see Rachen et al. (1995).

| source name | | | $\tau_{es}$ | $Y$ | $\alpha_s$ | $\alpha_h$ | $\epsilon_b$ [keV] | $\log[L^h_{17.7}/L_{15.5}]$ |
|---|---|---|---|---|---|---|---|---|
| PG 0026 +129 | | | 3.90 | 0.38 | $2.05 \pm 0.14$ | 0.86 | 0.17 | $-1.1$ |
| PG 0026 +129 | • | best-fit $n_H$ | 4.41 | 0.47 | $1.78 \pm 0.08$ | 0.86 | 0.63 | $-1.3$ |
| PG 0844 +349 | • | | 3.33 | 0.29 | $2.50 \pm 0.16$ | 1.0 | 0.30 | $-2.2$ |
| PG 0953 +414 | • | best-fit $n_H$ | 4.30 | 0.45 | $1.84 \pm 0.02$ | 1.0 | 3.49 | $-2.2$ |
| PG 1302 $-$102 | • | | 4.39 | 0.47 | $1.79 \pm 0.03$ | 0.5 | 1.45 | $-1.6$ |
| PG 1302 $-$102 | | best-fit $n_H$ | 4.09 | 0.41 | $1.96 \pm 0.04$ | 0.5 | 0.54 | $-1.4$ |
| PG 1411 +442 | • | | 3.33 | 0.29 | $2.50 \pm 0.04$ | 1.0 | 1.48 | $-3.3$ |
| PG 1634 +706 | • | | 3.58 | 0.33 | $2.30 \pm 0.14$ | 1.0 | 0.45 | $-2.1$ |
| PHL 1657 | • | best-fit UV | 4.32 | 0.45 | $1.83 \pm 0.03$ | 0.81 | 1.70 | $-2.0$ |
| PHL 1657 | | observed UV | 5.73 | 0.75 | $1.26 \pm 0.12$ | 0.81 | 2.24 | $-0.6$ |
| mean values | | | $4.0 \pm 0.5$ | $0.4 \pm 0.1$ | $2.1 \pm 0.3$ | — | $1.4 \pm 1.1$ | $-2.1 \pm 0.6$ |

Currently, there is no consensus about the origin of X-ray emission in AGN, and in particular about the soft X-ray excesses. It is our opinion that the key to understanding lies in the *link* between the UV-to-soft X-ray bump and the hard X-ray emission. Furthermore, it seems most natural to associate the big bump with a dense inflow and the hard X-rays with a tenuous outflow. Coupled accretion/ejection systems have been studied by numerous authors (e.g., Blandford & Payne 1982; Pelletier & Pudritz 1992, Ferreira & Pelletier 1993; Falcke et al. 1995).

In this Letter, our working hypothesis is that (i) hard X-rays are due to (anisotropic) nonthermal emission from a nuclear jet and (ii) soft X-rays are due to thermal comptonization of disk photons on the base of this jet where it emerges from the accretion disk.

## 2. The model

We consider the unsaturated comptonization of a thermal disk spectrum by the hot electrons from an optically thick base of a nuclear jet. This leads to a power law spectrum connecting the soft X-rays and the UV as a 'lever arm' (Shapiro et al. 1976). The 'pivot point' of this lever arm corresponds to the frequency where the maximum luminosity of the disk is emitted, i.e. at $\epsilon_{uv} \sim 20$ eV (Sun & Malkan 1989, Zheng 1991). The bolometric correction due to the soft X-ray excess is small. The geometry is that of a disk covered by a moving finite-height cylinder, but here we neglect the Doppler-shift introduced by the slow acceleration of the jet plasma lifting off the disk.

Firstly, let us express the kinetic luminosity of the jet in terms of the Eddington luminosity, i.e. $L_{kin} = \epsilon L_{Edd}$ with some $0 < \epsilon < 1$. We can then readily compute the optical depth of the cylinder for a line of sight perpendicular to its axis yielding $\tau_j = 0.4\epsilon/[\beta_j(\gamma_j - 1)r_{10}]$ where $\gamma_j$ is the jet Lorentz factor and $r_{10} = r_j/10r_G$ is the jet radius in units of the gravitational radius $r_G = 1.5 \cdot 10^{13} m_8$ cm, and where $m_8 = M/10^8 M_\odot$. Close to the base of the jet, we expect that the jet velocity is of the order of the escape velocity yielding $\tau_j \sim 3$ for $\beta = 0.3$, $\epsilon = 0.1$ and $r_{10} = 1$. In fact, if we extrapolate the implied particle density $n_p = \tau_j/\sigma_T r_j$ to the kpc-scale, we find a value of $n_p \sim 10^{-4}$ cm$^{-3}$ for $m_8 = 1$, in reasonable agreement with observations of the Cyg A Hot Spots (Roland et al. 1988).

In order to obtain theoretical spectra we have numerically integrated the stationary Kompaneet's equation following Sunyaev & Titarchuk (1985) using the discretization scheme of Chang & Cooper (1970). Titarchuk (1994) has shown that the diffusion approximation to comptonization is still valid even in the limit of optical depth $\tau_j \lesssim 1$, since the soft X-ray photons must have scattered several times in order to obtain their energy. The Kompaneet's equation reads

$$\frac{1}{x^2}\frac{d}{dx}x^4\left(\frac{dn}{dx} + n + n^2\right) = \gamma n - \gamma \frac{f(x)}{x^3} \quad , \qquad (1)$$

where $n(x)$ denotes the photon distribution, $x = h\nu/(m_e c^2)$ the dimensionless photon energy, $\gamma = \beta m_e c^2/(kT_e)$, and $\beta$ parametrizes the asymptotic photon escape probability, $P(t) \propto \beta \exp(-\beta t)$, for $t \to \infty$. For cylindrical geometry $\beta = 3\pi^2/(16[\tau_{es} + 2/3]^2)$ holds approximately. A more accurate treatment of the geometry yields $\beta = \frac{1}{2}\mu_1^2/\tau_{es}^2$ where $\mu_1$ is the first zero of the equation $-\mu J_1(\mu) + \frac{3}{2}\tau_{es}J_0(\mu) = 0$ and the $J_i$ denote Bessel functions. The resulting escape probability, however, does not deviate from the above approximation significantly. The source term $f(x) = f_0(x/x_{uv})^{-\alpha_d} \exp(-x/x_{max})$ describes the injection of photons from an accretion disk. The parameter $\gamma$ is related to the usual Compton-Y parameter by $Y = 4/\gamma$. For $Y < 1$, solutions to Eq.(2) are roughly of the form $S_\nu \propto \nu^{-\alpha_s} \exp[-h\nu/kT_e]$ where $\alpha_s = \sqrt{9/4 + 4/Y} - 3/2$.

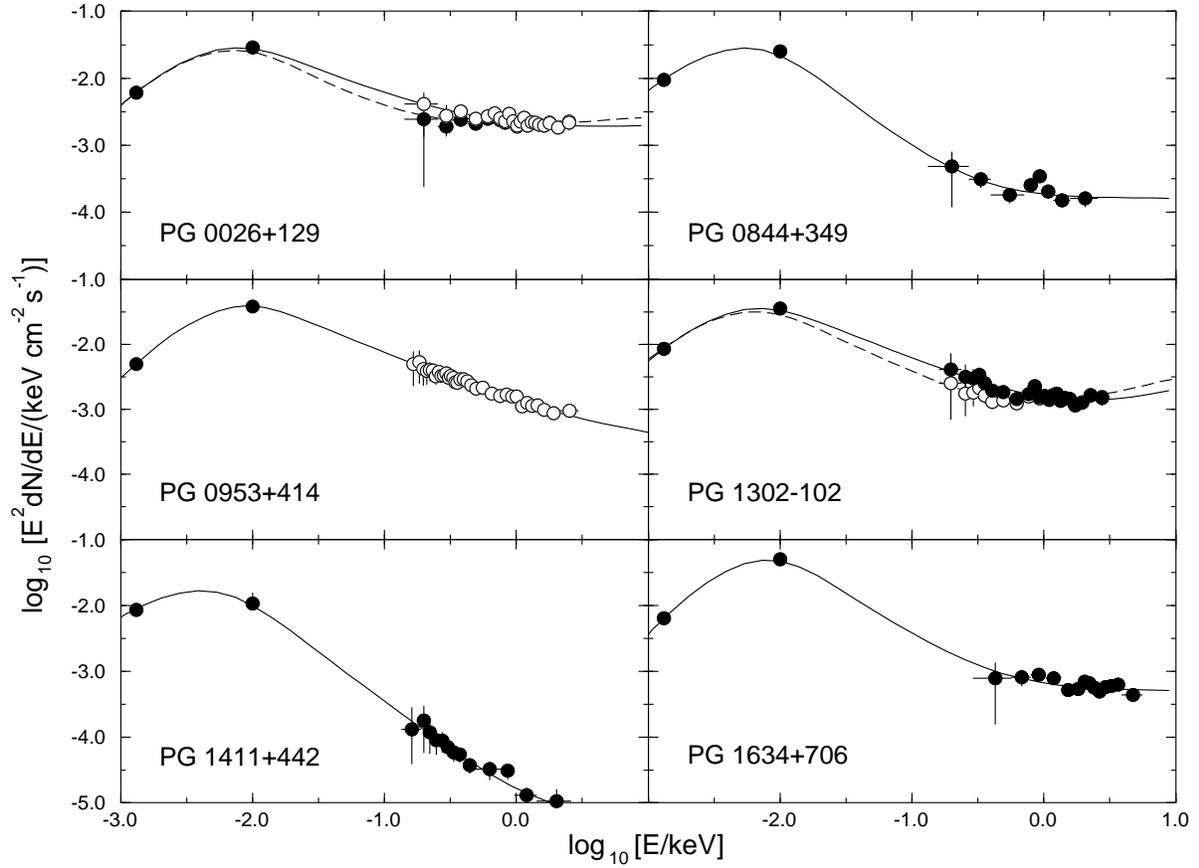

**Fig. 1.** Spectral energy distributions of the PG quasars in their source frame (cf. Elvis et al. 1994): The solid and dashed lines show the results of our model fits. De-reddened optical data points were taken from Sanders et al. (1989). UV continuum fluxes were extracted from Lanzetta et al. (1993) and de-reddened according to Seaton (1979). The soft X-ray data points are the unabsorbed $ROSAT$ PSPC fluxes binned for a two-power-law model with galactic $n_H$ (Rachen et al. 1995). The differences between the theoretical spectrum and the 2pl-model are negligible in the PSPC range. The x-axis error bars display the width of the photon bins, the y-axis error bars display statistical count rate errors and systematical errors due to the 1-$\sigma$ uncertainty in $n_H$. Open circles correspond to a free fit value of $n_H$.

We can make a simple argument on the expected value of the Compton-Y parameter, and hence predict a typical soft X-ray slope. Equating heating and cooling in the jet base yields

$$\frac{F_{uv}}{r_j} = \frac{4kT_e}{m_e c^2} \rho \kappa_{es} u_{rad}(1+\xi)c \quad , \qquad (2)$$

where $F_{uv}$ denotes the photon flux heating the jet base, $r_j$ the radius of the jet base and $\xi = u_B/u_{rad}$ the ratio of magnetic and radiation field energy densities, respectively. Since $u_{rad} = F_{uv}(1+\tau_j)/c$, the balance equation can be re-written yielding the Compton-Y parameter

$$Y = \frac{4kT_e}{m_e c^2} \tau_j [1+\tau_j] = \frac{1}{1+\xi} \quad . \qquad (3)$$

By definition $M_A^2 = u_{kin}/u_B$ and hence $L_{kin} = M_A^2 L_B$. At large radii $r \gg r_A$, the kinetic energy of a superalfvénic jet therefore exceeds the magnetic energy. The striking result from observations of radio jets is that $L_{disk} \sim L_j \sim L_{kin} + L_B$ (Rawlings & Saunders 1991, Celotti & Fabian 1993) implying that $\xi \sim [(1+M_A^2)\beta_j]^{-1}$ where $\beta_j$ denotes the dimensionless jet velocity. At the Alfvén radius we therefore have $\xi \sim [2\beta_j]^{-1} \sim 1$, which is also likely to be the case for $r < r_A$. Without a detailed jet model, it is difficult be more quantitative. It seems likely that even $\xi > 1$ is possible below the Alfvén radius, and their should be a trend for radio-quiets (presumably with slower jets) to produce somewhat larger $\xi$. Consequently, we expect $Y \lesssim 0.5$ or, equivalently, $\alpha_s \gtrsim 1.7$ (implying some infrared-to-optical cyclotron emission).

We applied the model to the spectra of six PG quasars and the weak bump quasar PHL 1657. A detailed description of the data can be found in separate paper by Rachen et al. (1995). In addition to a disk spectrum, we considered a hard X-ray power law of slope $\alpha_h = 1$ for radio-quiets and $\alpha_h =$ 0.5 for radio-louds, respectively. If hard X-ray slopes were available from EXOSAT or GINGA measurements, we took those (Tab.1). The slope of the soft X-ray spectrum $\alpha_{s,obs} = 2.1 \pm 0.3$ inferred from two-power-law fits to the data is in reasonable agreement with the theoretical value. The cross-over energy $\epsilon_b$, where the hard and soft X-ray power laws intersect, is constrained by the energetics of the coupled jet/disk system. Let us firstly define the efficiency $\delta$ for converting kinetic power into hard X-ray luminosity $L_h = \Lambda \nu F_\nu^{(h)}(\epsilon) = 0.1\delta_{0.1}L_{uv}$, where $\Lambda = \ln[\epsilon_{max}/\epsilon_{min}] \simeq 5$ and $\delta = 0.1\delta_{0.1}$. A value around $L_h/L_{uv} = 0.1$ corresponds to a typical ratio of the integrated observed hard X-ray luminosity to the disk luminosity (Elvis et al. 1994), for our sample the corresponding value obtained from Tab. 1 is $\sim 0.04$. Adopting a soft X-ray power law $\nu F_\nu^{(s)}(\epsilon) = L_{uv}(\epsilon/\epsilon_{uv})^{1-\alpha_s}$ and equating the $\nu F_\nu$ values for the two X-ray components, we expect the crossover energy to be close to the value $\epsilon_b \sim 1/\delta_{0.1}$ keV ($\epsilon_{uv} = 20$ eV, $\alpha_s = 2$). Thus, a high radiative efficiency of the jet $\delta > 0.1$ would produce a cross-over energy well below keV, whereas a low efficiency $\delta < 0.1$ (or large anisotropy, e.g. Mannheim 1995) would lead to the complete dominance of the steep comptonized disk spectrum in the keV range. The observed cross-over energy $\epsilon_b = 1.4 \pm 1.1$ keV is in agreement with radiative efficiencies of order 10%. It is interesting to note that PG 1411+442, which is the steepest source in our sample corresponding to the jet with lowest radiative efficiency, is a BAL quasar. Possibly, the jet is too slow in this source to operate as an efficient nonthermal radiator, and the clumpy breeze is seen as the BAL gas.

## 3. Conclusions

We predict that the UV bump spectrum of quasars extends into the soft X-ray range with a steep power law. For radio-louds, the spectral index is $\alpha_s \simeq 1.7$ ($S_\nu \propto \nu^{-\alpha}$), and for radio-quiets it should be somewhat steeper $\alpha_s \gtrsim 1.7$. The prediction is based on a simple scenario in which all quasars, radio-quiets and radio-louds, are assumed to have jets close to their accretion disks. Soft X-rays are produced via unsaturated Comptonization of UV photons from the disk in the (marginally) optically thick jet base. Nonthermal hard X-rays are produced further downstream. The scenario requires that (i) the jet kinetic luminosity is always of the same order as the UV luminosity and (ii) that the nonthermal radiative efficiency is $\sim 10\%$. In radio-quiets, the jets should have speeds of $\sim 0.3c$ and a low Mach number leading to rapid decollimation and turbulent mass entrainment within the central parsec. If the scenario is correct, the kinetic power of nuclear outflows inferred from optical emission line asymmetries must be a significant fraction of the electromagnetic luminosity. The recent discovery of a $\sim 50$ keV turnover in low-luminosity Seyferts is not in contradiction with a nonthermal nature of the hard X-ray source. The observed X-ray hump could represent a strong reflection component superimposed on a weak nonthermal component, as predicted by anisotropic cascade models (Mannheim 1995). The proposed origin of the soft X-rays could be further tested by observations of X-ray spectral variability. Injection of magnetic flares at the jet base increases the Compton $Y$-parameter for a dynamical time $\Delta t_s \approx r_j/v_j \lesssim 10(L_{uv}/10^{46}$ erg s$^{-1})$ days (adopting $L_{uv}/L_{edd} = 0.1$). The comptonized soft X-ray spectra should vary on these time scales according to a "pivoting" where the pivot point lies in the UV. The hard X-rays coming from further downstream are not affected immediately. Thus, hard X-ray observations would diagnose steepening with increasing flux, whereas soft X-ray observations would see a flattening with increasing flux.

*Acknowledgements.* This research was partly supported by DARA grant FKZ 50 OR 9202.